\documentclass[12pt]{iopart}

\usepackage{epsfig}
\usepackage{graphicx,psfrag,xspace}

\begin{document}

\title[]{Time to reach the maximum for a random acceleration process}

\author{Satya N. Majumdar, Alberto Rosso}
\address{CNRS - Universit\'e Paris-Sud, LPTMS, UMR8626 - B\^at.~100, 91405 Orsay Cedex, France}

\author{Andrea Zoia}
\address{CEA/Saclay, DEN/DM2S/SERMA/LTSD, B\^at.~454, 91191 Gif-sur-Yvette Cedex, France}
\ead{andrea.zoia@cea.fr}

\begin{abstract}
We study the random acceleration model, which is perhaps one of the simplest, yet nontrivial, non-Markov stochastic processes, and is key to many applications. For this non-Markov process, we present exact analytical results for the probability density $p(t_m|T)$ of the time $t_m$ at which the process reaches its maximum, within a fixed time interval $[0,T]$. We study two different boundary conditions, which correspond to the process representing respectively (i) the integral of a Brownian bridge and (ii) the integral of a free Brownian motion. Our analytical results are also verified by numerical simulations. 

\end{abstract}

\maketitle

\section{Introduction}

Consider a general stochastic process $x(t)$, starting from $x(0)=0$, over a 
fixed time interval $[0,T]$. Let $t_m$ denote the time at which the process 
achieves its maximum value $x_m$ during the interval $[0,T]$ (see Fig. 1). 
Clearly, $t_m$ 
is a random variable that fluctuates from one realization of the process to 
another. Our main goal is to investigate the probability density function 
(pdf) $p(t_m|T)$ of the stochastic times $t_m$, given the interval length $T$ 
and the underlying stochastic process $x(t)$.

\begin{figure}[t]
   \centerline{ \epsfclipon \epsfxsize=7.0cm
\epsfbox{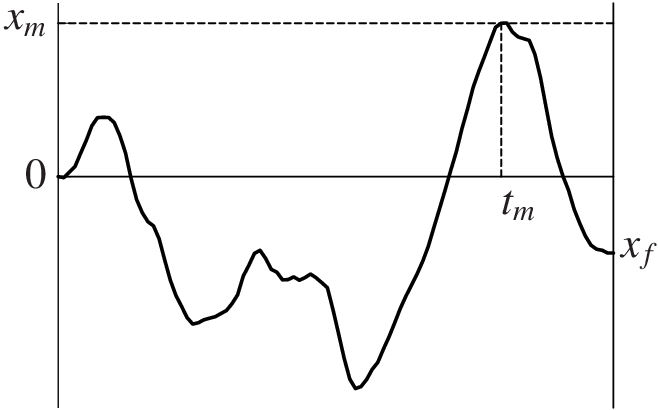} }
   \caption{A realization of a random acceleration process reaching its maximum $x_m$ at $t_m$.}
   \label{fig1}
\end{figure}

This question naturally rises in a variety of fields. For instance, in the 
context of the queueing theory, the stochastic process $x(t)$ may represent 
the length of a queue at time $t$, and one would like to know $t_m$, i.e., the 
time at which the queue length is maximum over a given time span. In the 
context of finance, $x(t)$ may represent the price of a stock, and a trader is 
evidently interested in $t_m$, i.e., the time at which the stock price is at 
its highest.

Perhaps, one of the simplest examples is provided by the underlying stochastic process $x(t)$ being an ordinary Brownian motion
\begin{equation}
\frac{dx}{dt}= \eta(t),
\label{brown1}
\end{equation}
where $\eta(t)$ is a Gaussian white noise with zero mean 
$\langle \eta(t)\rangle=0$, and a delta correlator 
$\langle \eta(t)\eta(t')\rangle= 2D\delta(t-t')$. In this case, the probabiliy 
density $p(t_m|T)$ is independent of $D$ and is well known~\cite{Levy}
\begin{equation}
p(t_m|T)= \frac{1}{\pi\, \sqrt{t_m(T-t_m)}};\quad 0\le t_m\le T.
\label{levy1}
\end{equation}
This curve is {\em symmetric} around the mid-point $t_m=T/2$, 
has a `$U$ shape' with minimum at $t_m=T/2$, and 
diverges at the two end points $t_m=0$ and $t_m=T$. 
The cumulative distribution 
${\rm Prob}(t_m\le x|T)= 
\frac{2}{\pi}{\arcsin}\left[\sqrt{t_m/T}\right]$ 
is known as one of Levy's celebrated `arcsine laws'~\cite{Levy}.

For a Brownian bridge, i.e., a Brownian motion constrained to return to $0$ at 
$T$, i.e., $x(T)=0$, the corresponding pdf is known to be uniform: $p(t_m|T)= 1/T$ 
for $0\le t_m\le T$~\cite{Feller}.

Motivated principally by the applications in the queueing theory and finance, 
the pdf $p(t_m|T)$ has been computed explicitly for a variety of other 
`constrained' Brownian motions by using suitably adapted path integral 
methods. These examples include Brownian excursion~\cite{Yor}, Brownian 
meander~\cite{Yor}, reflected Brownian bridge~\cite{Yor}, Brownian motion till 
its first-passage time~\cite{RM}, and Brownian motion with a drift~\cite{MB}. 
Some of these exact results were then reobtained via a functional 
renormalization group method~\cite{SL}. In Ref.~\cite{SL}, the authors 
computed $p(t_m|T)$ also for Bessel processes and for continuous-time random 
walks. Curiously, the pdf $p(t_m|T)$ also appeared recently as an important 
input in the calculation of the area enclosed by the convex hull of a planar 
Brownian motion of duration $T$~\cite{hull1,hull2}, which displays an 
interesting application in ecology. In addition, we have recently shown that 
$p(t_m|T)$ also appears in the computation of the disorder-averaged 
equilibrium distribution of a particle moving in a random self-affine 
potential~\cite{MRZ}.

The results for $p(t_m|T)$ mentioned above have been obtained for Brownian motion and its variants, which are all Markov processes. The purpose of this paper is to go beyond Markov processes and present an exact result of $p(t_m|T)$ for a non-Markov process. The non-Markov process that we study here is the well-known random acceleration process, which evolves via
\begin{equation}
\ddot{x}(t)=\eta(t),
\end{equation}
$\eta(t)$ being a Gaussian white noise with $\langle \eta(t)\rangle =0$, as before, and $\langle \eta(t)\eta(t')\rangle = 2D\delta(t-t')$. For simplicity, we will choose $D=1$ subsequently. The process starts at $x(0)=0$ with $v(0)=\dot{x}(t)=0$, and evolves until the time $T$. We denote by $x_f=x(T)$ and $v_f=v(T)$ the final position and velocity of the process, respectively (see Fig. 1). The random acceleration process in the single variable $x(t)$ is non-Markovian, whereas the position-velocity process $\left\lbrace x, v\right\rbrace $ is Markovian and does not depend on the past history.

The random acceleration process, being among the simplest non-Markovian processes, has been intensely studied both in Physics and Mathematics literature. In Physics, for instance, it appears in the continuum description of the equilibrium Boltzmann weight of a semiflexible polymer chain with non-zero bending energy~\cite{Burkhardt}. It also describes the steady state profile of a $(1+1)$-dimensional Gaussian interface~\cite{MB1} with dynamical exponent $z=4$, the continuum version of the Golubovic-Bruinsma-Das Sarma-Tamborenea model~\cite{GBDT}. This process also appears in the description of the statistical properties of the Burgers equation with Brownian initial velocity~\cite{PV}. The first-passage and a variety of other related properties of the random acceleration model are highly nontrivial and have been studied extensively over the last few decades~\cite{McKean,Goldman,MW,Burkhardt,Sinai,MB2,MB1,Lachal,GL,Burkhardt2}.

Thus, in addition to being relevant in many applications, the random acceleration model represents a simple, yet nontrivial, non-Markov process where one can try to compute observables of physical interest. Recent studies have concerned the distribution of extreme observables associated with this process, notably the global maximum $x_m$ itself over the interval $[0,T]$~\cite{Burkhardt,Gyorgy,Burkhardt3}. In this paper, we focus instead on the time $t_m$ at which the global maximum $x_m$ occurs in $[0,T]$. Actually, we show in this paper that $p(t_m|T)$ can be computed explicitly.

Let us first summarize our main results. Since the only time scale in the problem is $T$, it is evident that $p(t_m|T)$, normalized to unity over $0\le t_m\le T$, has the scaling form
\begin{equation}
p(t_m|T)= \frac{1}{T}\, p\left(\frac{t_m}{T}\right),
\label{scaling1}
\end{equation}
where the scaling function $p(z)$, defined over $0\le z\le 1$, satisfies
the normalization condition: $\int_0^1 p(z)dz=1$.
We will consider 
two different boundary conditions, detailed below, for which we
are able to compute $p(z)$ explicitly.

\begin{figure}[t]
   \centerline{ \epsfclipon \epsfxsize=7.0cm
\epsfbox{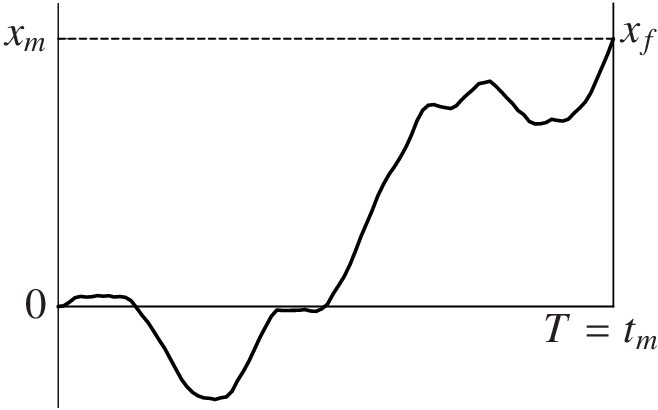} }
   \caption{A realization of a random acceleration process reaching its maximum $x_m$ at $t_m$. In this case, the maximum occurs at the end of the time interval $[0,T]$.}
   \label{fig2}
\end{figure}

\subsection{Integral of a Brownian Bridge}

When the final velocity vanishes, $v_f=0$, the process can be interpreted as the integral of a Brownian Bridge. In this case, we will show that
\begin{equation}
p(z)=\frac{\Gamma(1/2)}{\Gamma^2(1/4)}\frac{1}{\left[z ( 1-z )  \right]^{3/4} },
\label{resultbridge}
\end{equation}
which is evidently {\em symmetric} around the mid-point $z=1/2$ and 
diverges at the two end-points as $z^{-3/4}$ and $(1-z)^{-3/4}$, respectively.
It is also useful to consider the cumulative distribution
$P(z)= \int_0^{z} p(z')\, dz'$, which reads
\begin{equation}
P(z)= \frac{\Gamma(1/2)}{\Gamma^2(1/4)}\, 
B_z\left(\frac{1}{4},\frac{1}{4}\right).
\label{cum_bridge}
\end{equation}
Here $B_z(p,q)=\int_0^z x^{p-1}\,(1-x)^{q-1}\,dx$ is the incomplete
Beta function~\cite{GR}. A plot of $P(z)$ is shown in Fig. 4, where
it is also compared to direct numerical solutions, to an
excellent agreement.

\subsection{Integral of a free Brownian Motion}

When $v_f$ is arbitrary, the process can be interpreted as the integral of a free Brownian motion. In this case, we obtain the following exact result for the normalized pdf
\begin{equation}
p(z)=C\,\delta(z-1)+\frac{(1-C)}{\pi \sqrt{2}} z^{-3/4}\,(1-z)^{-1/4},
\label{result2}
\end{equation}
where the constant $C$ has the exact value
\begin{equation}
C=1-\sqrt{\frac{3}{8}}=0.387628\ldots
\label{resultC}
\end{equation}
The density is thus {\em asymmetric} around the mid-point 
$t_m=T/2$ (i.e., $z=1/2$), and the maximum may either occur at some time {\em 
strictly} shorter 
than 
$T$, namely $t_m<T$ (i.e., $z<1$), or with a finite nonvanishing probability 
$C=0.387628..$ at the end point of the interval 
$t_m=T$ (or equivalently $z=1$). The representative trajectories
for these two situations are shown respectively in Fig. 1 and Fig. 2.
In other words, roughly $38.67\%$ of all trajectories, 
starting at $x(0)=0$ and $v(0)=0$, achieve their 
maximum only at the end of the interval $[0,T]$. 
The corresponding cumulative distribution $P(z)=\int_0^{z} p(z')\, dz'$
is given by
\begin{equation}
P(z)= C\, \theta(z-1) + \frac{(1-C)}{\pi \sqrt{2}}\, B_z\left(\frac{1}{4},\frac{3}{4}\right),
\label{cum_free}
\end{equation}
where $\theta(z-1)$ is zero for $z<1$ and is equal to $1$ for $z=1$,
i.e, $P(z)$ exhibits a discontinuous jump at $z=1$ from 
$1-C=\sqrt{3/8}=0.612372\dots$ to $1$. A plot of $P(z)$ is provided
in Fig. 5, where it is also compared to direct simulation results: we find an excellent agreement between analytical and numerical
results.

The paper is organized as follows. In Section 2, we outline the main ideas 
used for the exact derivation of $p(t_m|T)$ via path decomposition techniques.
In Subsection 2.3, we also compare our analytical predictions
with the results obtained via Monte Carlo simulations. 
A concluding Section 3 presents a summary and raises some open questions. The details of the 
calculations 
are left to four Appendices, as they involve rather cumbersome multiple 
integrations of Airy functions.

\section{Calculations}

The basic ingredient in our computation is the propagator $Z_+(x_1,v_1;x_0,v_0,t)$, i.e., the probability that the process $x(t)$, starting at $x_0>0$ with velocity $v_0$, reaches the point $x_1$ with velocity $v_1$ at time $t$, without ever crossing the origin $x=0$ during this time. The Laplace transform of the propagator has been explicitly computed by Burkhardt and is recalled in~\ref{a:propagator}. We show that the location of the maximum can be expressed in terms of the propagator $Z_+$. Then, we make use of the known results about $Z_+$ in Laplace space to derive a closed form expression for $p(t_m|T)$.

The velocity $v(t)$ of a random acceleration process is a Brownian motion, thus $v(t)$ is continuous everywhere, which implies that $x(t)$ is continuous and differentiable everywhere. An important consequence is that the global maximum $x_m$ can lie either inside the interval $(0,T)$, with a velocity $v_m=0$ (Fig. 1), or at the boundary $t_m=T$, with a velocity $v_m=v_f >0$ (Fig. 2).

With reference to Fig. 1, the global maximum of the process starting at $\left\lbrace x_0=v_0=0\right\rbrace$ and ending at $\left\lbrace x_f, v_f\right\rbrace $ lies inside the interval $(0,T)$. It is useful to decompose the total path in two portions: a first path from $\left\lbrace x(0)=0, v(0)=0 \right\rbrace $ to $\left\lbrace x_m, v_m=0 \right\rbrace $ in a time $t_m$, and a second path from $\left\lbrace x_m, v_m=0 \right\rbrace $ to $\left\lbrace x_f, v_f \right\rbrace $ in a time $T-t_m$ (see Fig. 3a). Remark that $x_m$ represents the maximum of both paths. Denote by $\mbox{P}[x_m, v_m=0; x(0)=v(0)=0,t_m]$ the probability of the first path and by $\mbox{P}[x_f, v_f;x_m, v_m=0, T-t_m]$ the probability of the second path. With respect to the variables $\left\lbrace x, v \right\rbrace $, the process is Markovian, and the probability of the entire path is given by the product
\begin{equation}
\mbox{P}[x_m, v_m=0; x(0)=v(0)=0,t_m] \times \mbox{P}[x_f, v_f;x_m, v_m=0, T-t_m].
\label{twopaths}
\end{equation}

\begin{figure}[t]
   \centerline{ \epsfclipon \epsfxsize=7.0cm
\epsfbox{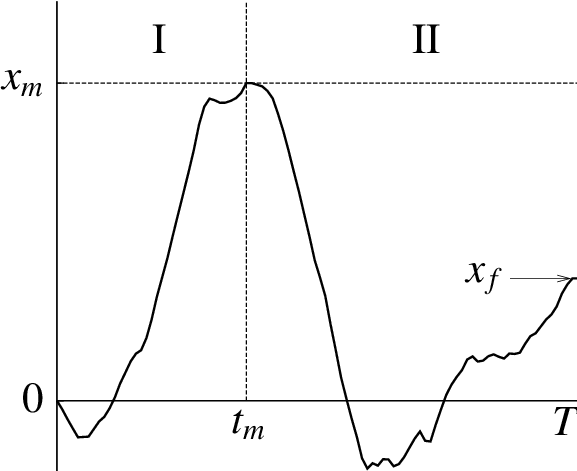} \, \epsfbox{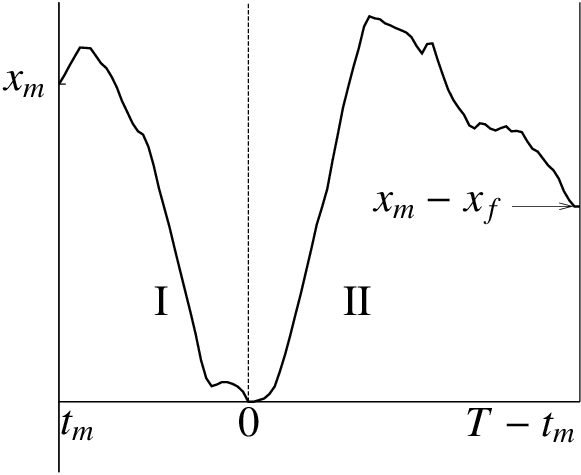}}
   \caption{a) Left. Decomposition of a realization of the random acceleration process into two paths, from $0$ to $t_m$ (I) and from $t_m$ to $T$ (II). b) Right. Illustration of the change of variables $\tilde{x}=x_m -x$ and $\tilde{t}=t_m -t$ for (I) and $\tilde{x}=x_m -x$ and ${\tilde{t}}=t-t_m$ for (II).}
   \label{fig3}
\end{figure}

We introduce the change of variables $\tilde{x}=x_m -x$, $\tilde{t}=t_m -t$ (which implies $\tilde{v}=v$) for the first path (see Fig. 3b). Therefore, we can rewrite
\begin{equation}
\mbox{P}[x_m, v_m=0; x(0)=v(0)=0,t_m] =Z_+(x_m,0;0,0,t_m).
\end{equation}

As for the second path, the change of variables $\tilde{x}=x_m -x$, ${\tilde{t}}=t-t_m$ (see Fig. 3b) gives
\begin{equation}
\mbox{P}[x_f, v_f;x_m, v_m=0, T-t_m] =Z_+(\tilde{x}_f= x_m-x_f,-v_f;0,0,T-t_m).
\end{equation}

When $t_m=T$, as illustrated in Fig. $2$, an analogous argument, with a change of variables $\tilde{x}=x_m -x$ and $\tilde{t}=T -t$ (which implies $\tilde{v}=v$), allows rewriting the probability of such a path as $Z_+(x_m,0;0,v_f,T)$.

\subsection{Integral of a Brownian Bridge}

In this case, $v_f=0$, and the location of the maximum lies inside the interval, as in Fig. 1. It follows that $p(t_m|T)$ can be obtained by integrating the two paths of Eq.~(\ref{twopaths}) over $x_m$ and $\tilde{x}_f$, with $v_f=0$
\begin{eqnarray}
\int_0^\infty d x_m \int_{-\infty}^{x_m} d x_f \mbox{P}[x_m, 0;0,0,t_m]  \mbox{P}[x_f, 0;x_m, 0, T-t_m]= \nonumber \\
\int_0^\infty d x_m Z_+(x_m,0;0,0,t_m) \int_0^\infty d \tilde{x}_f Z_+(\tilde{x}_f,0;0,0,T-t_m) = \nonumber \\    
I_2(0,t_m) I_2(0,T-t_m).
\label{prop2}
\end{eqnarray}
We define
\begin{equation}
I_2(\epsilon,t)= \int_0^{\infty} dx \, Z_{+}(x,0;\epsilon,0,t),
\label{I2def}
\end{equation}
which denotes the probability that the process, starting at the initial position $\epsilon \ge 0$ at $t=0$, with initial velocity $v_0=0$, remains positive up to time $t$, with a vanishing final velocity (the final position being arbitrary).

The integral $I_2(\epsilon,t)$ can be computed exactly and it turns out that, in the $\epsilon \to 0$ limit, this integral vanishes, as apparent from Eq.~(\ref{I2final}). This is not surprising: indeed, if the particle starts at the origin with vanishing velocity, it can not survive up to a finite time $t$, since it will cross the boundary almost immediately. This is due to the continuous nature of the Brownian velocity. Therefore, if one puts $\epsilon=0$ straightaway, the rhs of Eq.~(\ref{prop2}) vanishes. The reason for this is clear: the lhs of Eq.~(\ref{prop2}) represents the probability that the maximum lies in $[t_m,t_m+dt_m]$, i.e., $p(t_m|t)dt_m$, and hence is proportional to the small time increment $dt_m$. Setting $\epsilon\to 0$ essentially implies $dt_m=0$, which therefore gives $0$ on the rhs. To extract the nonzero {\em probability density} $p(t_m|t)$, one therefore needs to keep a finite $\epsilon$ (therefore a finite $dt_m$) in $I_2(\epsilon,t)$ on the rhs of Eq.~(\ref{prop2}). Then, finally one uses the following limiting procedure that gives a finite answer
\begin{equation}
p(t_m|T)=\mathop {\lim }
\limits_{\epsilon \to 0 }
\frac{I_2(\epsilon,t_m) 
I_2(\epsilon,T-t_m)}{\int_0^T d t_m I_2(\epsilon,t_m) 
I_2(\epsilon,T-t_m)}= 
\frac{\Gamma(1/2)}{\Gamma^2(1/4)} 
\frac{\sqrt{T}}{\left[t_m ( T-t_m)  \right]^{3/4} }.
\end{equation}
We note that this type of regularization procedure has been adopted before in computing the distribution of the area under a Brownian excursion~\cite{Airy} and, more recently, in the context of computing $p(t_m|T)$ for several constrained Brownian motions, including the Brownian excursion~\cite{Yor}. 

\subsection{Integral of a free Brownian Motion}

When the final velocity $v_f$ of the particle is arbitrary, there is a finite probability that the global maximum occurs at the end of the observation time window $[0,T]$. This means that the probability density $p(t_m|T)$ includes a delta function $C\delta(t_m-T)$ (with a nonzero weight $C$ ). This is in addition to a non-delta function part with a nonzero support (and a total weight $(1-C)$) over the full interval $t_m\in [0,T]$.

Let us first compute the delta-function component in $p(t_m|T)$ at $t_m=T$. This
contribution comes from paths that start at the origin with zero velocity
and end up at $x_m$ (with $x_m$ being the maximum) in a small
time window $t_m\in [T-\delta,T]$, where $\delta\to 0$ and 
$x_m$ is free to take any positive value. Following the notations
and the change of variables discussed earlier, one can write
the net probability of such paths as 
\begin{equation}
\lim_{\delta \to 0}\int_{T-\delta}^{T}p(t_m|T)\, dt_m = \int_0^\infty 
\int_0^\infty d x_m d v_f Z_+(x_m,0;0,v_f,T)=I_1(T).
\end{equation}
We compute this integral $I_1(T)$  explicitly in ~\ref{I1},
and it turns out to be a constant indepedent of $T$,
$I_1(T)= C = 1-\sqrt{3/8}$. 
Thus, it follows that the delta-function component
of the probability density is $p(t_m|T)= C \delta(t_m-T)$. 

We next focus on the non-delta function part, where the maximum
occurs at some time $t_m$ well inside the interval $[0,T]$.
The calculation is performed along the lines of the integral of a  Brownian 
Bridge. The integral of the two paths of Eq.~(\ref{twopaths}) now writes
\begin{eqnarray}
  \int_0^\infty d x_m Z_+(x_m,0;0,0,t_m)  \int_{-\infty}^{+\infty} d \tilde{v}_f \int_0^\infty d \tilde{x}_f Z_+(\tilde{x}_f,- v_f;0,0,T-t_m) \nonumber \\
= I_2(0,t_m) I_3(0,T-t_m),  
\label{prop3}
 \end{eqnarray}
where
\begin{equation}
I_3(\epsilon,t)= \int_{-\infty}^{+\infty} dv\, \int_0^{\infty} dx\,
Z_{+}(x,v;\epsilon,0,t).
\label{I3def}
\end{equation} 

The integral $I_3(\epsilon, t)$ 
is computed in Eq.~(\ref{I3}) and, for reasons explained already, it
vanishes as $\epsilon\to 0$. Using the 
regularization mentioned before with $x_0=\epsilon$ and taking into
account the full normalization, $\int_{t_m}^{T} p(t_m|T)\,dt_m=1$,
we can then write
\begin{equation}
p(t_m|T)= C\delta(t_m-T)+(1-C) \mathop {\lim }\limits_{\epsilon \to 0 
}\frac{I_2(\epsilon,t_m) I_3(\epsilon,T-t_m)}{\int_0^T d t_m 
I_2(\epsilon,t_m) I_3(\epsilon,T-t_m)  } 
\end{equation}
This thus gives the result of Eq.~(\ref{result2}).

\subsection{Numerical simulations}

To verify our main theoretical predictions in the two cases, namely,
(i) the integral of a Brownian bridge in Eq. (\ref{cum_bridge})
and (ii) the integral of a free Brownian motion in Eq. (\ref{cum_free}),
we have also performed Monte Carlo simulations of the two processes.
In both cases, we
simulated $10^4$ realizations, with $T=1$ and an integration step $10^{-5}$.
In Fig. 4 and 5, for the two cases (i) and (ii) respectively, the 
theoretical and numerical results are compared.
The circles represent the simulation
results and the solid lines represent the analytical formulae,
respectively in Eq. (\ref{cum_bridge}) (Fig. 4) and
Eq. (\ref{cum_free}) (Fig. 5).
The agreement between simulations and analytical curves is excellent.

\begin{figure}[t]
   \centerline{ \epsfclipon \epsfxsize=8.0cm
\epsfbox{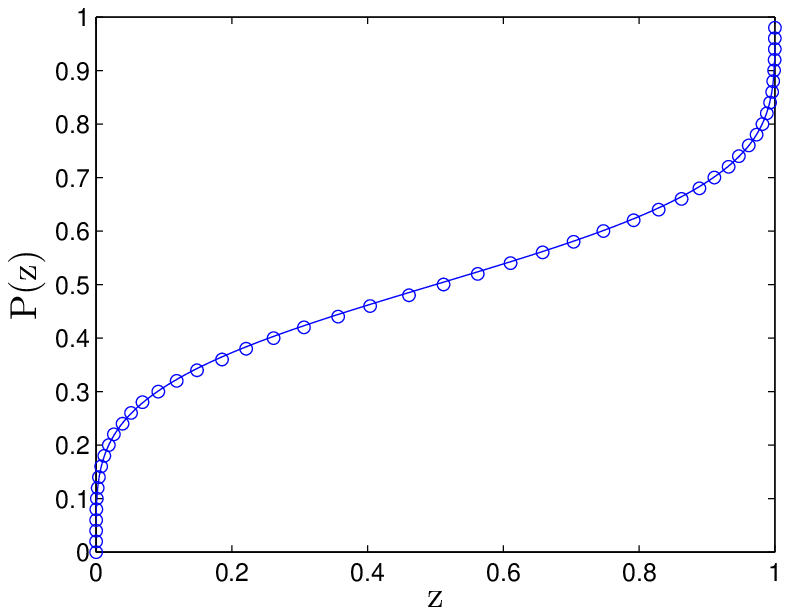} }
   \caption{Simulation results for the cumulative distribution
$P(z)=\int_0^z p(z')\, dz'$ (circles) as 
compared to the analytical 
formula in Eq.~(\ref{cum_bridge}) (solid 
line), 
for the integral of a Brownian Bridge.}
   \label{fig4}
\end{figure}

\begin{figure}[t]
   \centerline{ \epsfclipon \epsfxsize=8.0cm
\epsfbox{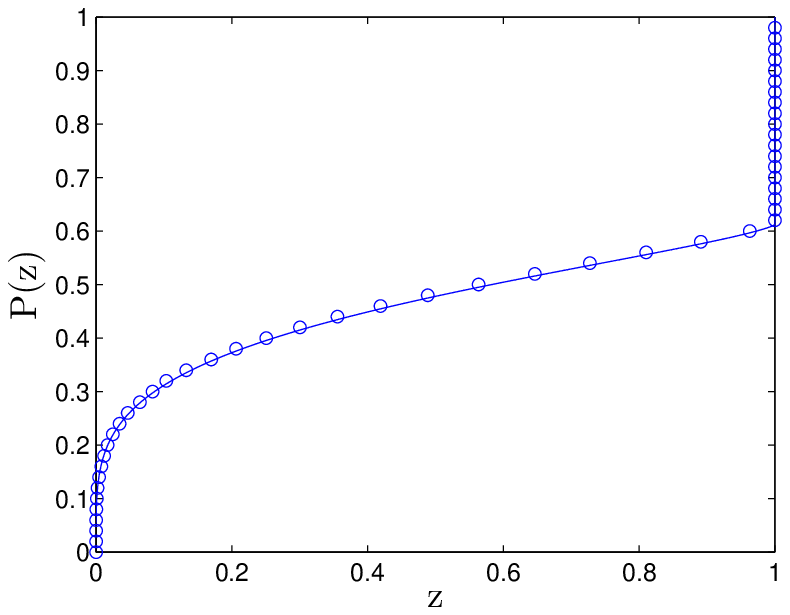} }
   \caption{Simulation results for the cumulative
distribution $P(z)$ (circles) as compared to the 
analytical 
formula in Eq.~(\ref{cum_free}) (solid line), for the integral of a free 
Brownian motion.}
   \label{fig5}
\end{figure}

\section{Summary and conclusions}

In this paper we have studied two non-Markov stochastic processes: 
$(i)$ the integral of a Brownian bridge up to time $T$ and $(ii)$
the integral of a Brownian motion up to $T$ - i.e., the so called
random acceleration processes. For both processes, we have
computed exactly the probability density $p(t_m|T)$
of the time $t_m$ at which the process achieves its maximum
when observed over a time window $[0,T]$.
In the former case, the density $p(t_m|T)$ is symmetric
around the midpoint $t_m=T/2$, with power law divergences
at the end points $t_m=0$ and $t_m=T$. In contrast, in
the latter case, $p(t_m|T)$ is asymmetric around the
midpoint $t_m=T/2$ and, in addition, has an unusual
delta-function contribution at the end point $t_m=T$.
The exact results are given respectively in Eqs.~(\ref{resultbridge})
and~(\ref{result2}). Our analytical findings are in excellent agreement
with numerical Monte Carlo simulations (see Figs. 4 and 5). 

Note that, even though our final results look rather simple, their derivations,
starting from the basic propagator, are nontrivial
and involve rather delicate mathematical manipulations of
multiple integrals, which we have tried to present systematically
in a stepwise fashion in the Appendices. Hopefully, some of these
details might be also useful in other problems involving the integrals
of a Brownian motion.

The exact result for $p(t_m|T)$ that we obtain in this paper
is directly relevant for a particle diffusing in 
a one dimensional random potential $V(x)$.
In a finite box of size $T$, the particle finally
thermalizes with an equilibrium Boltzmann
density, $p_{\rm eq}(x)= e^{-\beta V(x)}/Z$,
where $\beta$ is the inverse temperature and
$Z=\int_0^T  e^{-\beta V(x)}\, dx$ is the partition function.
In the low temperature limit, by taking the average
over disorder, one can show~\cite{MRZ} that
$\overline{p_{\rm eq}(x)}$ is precisely equal
to the probability density $p(t_m=x|T)$ that the potential
$V(x)$, viewed as a stochastic process in space (where the space
plays the role of `time'), has its minimum (or equivalently
its maximum~\cite{foot})
at $t_m=x$. 
For example, for the well known Sinai model, where $V(x)$ itself
is a free Brownian trajectory, $\overline{p_{\rm eq}(x)}=1/[\pi 
\sqrt{x(T-x)}]$, the same as the arcsine law. Our finding for
$p(t_m|T)$ then generalizes the
result for $\overline{p_{\rm eq}(x)}$ to the case where
the potential $V(x)$ represents the trajectory of
a random acceleration process.
 
Let us end with a curious observation. For the free Brownian
motion, Levy's arcsine law appears in the distribution of two
different observables: (a) in $p(t_m|T)$, i.e., the location
of the maximum, and (b) in $p(t_{\rm occup}|T)$, where
$t_{\rm occup}=\int_0^T \theta\left[x(\tau)\right]\, d\tau$
represents the occupation time, i.e., the time the process
spends on the positive half axis within the interval $[0,T]$.
One may then ask if these two observables share the same
distribution also for the random acceleration process.
While we are able to compute $p(t_m|T)$ explicitly
in this case, the computation of $p(t_{\rm occup}|T)$
seems harder in the random acceleration case.
Our numerical evidence shows
that, unlike in the Brownian case, the two distributions 
are different in the random acceleration model. 
Therefore, computing the occupation time distribution
$p(t_{\rm occup}|T)$ for the random acceleration model
remains a challenging open  problem.

\appendix

\section{Propagator with an absorbing boundary}
\label{a:propagator}

In the absence of boundaries, the free propagator for the random acceleration process reads~\cite{Burkhardt}
\begin{equation}
Z_0(x,v;x_0,v_0,t)=\frac{\sqrt{3}}{2\pi t^2}\exp\left\lbrace  
-\frac{3}{t^{3}} \left[  \left( x-x_0-vt\right) \left( x-x_0-v_0t\right) +\frac{t^2}{3}\left( v-v_0\right)^2\right] \right\rbrace .
\label{z0}
\end{equation}
In presence of the absorbing boundary at $x=0$, the propagator reads~\cite{Burkhardt}
\begin{equation}
Z_+(x,v;x_0,v_0,t)= Z_0(x,v;x_0,v_0,t) + Z_1(x,v;x_0,v_0,t),
\label{z+}
\end{equation}
where the Laplace transform of $Z_1(x,v;x_0,v_0,t)$, i.e.,
\begin{equation}
\tilde{Z}_1(x,v;x_0,v_0,s)=\int dt e^{-st} Z_1(x,v;x_0,v_0,t),
\label{z1}
\end{equation}
has a rather complicated expression~\cite{Burkhardt}
\begin{eqnarray}
 \tilde{Z}_1(x,v;x_0,v_0,s)= -\frac{1}{2 \pi} \int^{\infty}_{0} 
\int^{\infty}_{0}dG   dF \mbox{Ai}(s/G^{2/3}-vG^{1/3})\nonumber \\
\mbox{Ai}(s/F^{2/3}+ v_0 F^{1/3})\frac{\exp\left\lbrace - \left[ F x_0+ G x +\frac{2}{3} s^{3/2} \left( F^{-1}+G^{-1}\right) \right]  \right\rbrace }{(FG)^{1/6} (F+G)},
\label{z1l}
\end{eqnarray}
$\mbox{Ai}(x)$ being the Airy function~\cite{AS}. 
In subsequent calculations, we will also use the following amazing 
identity~\cite{MW,Burkhardt}
\begin{equation}
\frac{1}{2 \pi} \int^{\infty}_{0} \frac{dG}{F+G}\, G^{-1/6}\, \exp\left[-\frac{2}{3}\left(\frac{s^{3/2}}{F}+\frac{s^{3/2}}{G}\right)\right]\mbox{Ai}\left(\frac{s}{G^{2/3}}\right)= F^{-1/6}\, \mbox{Ai}\left(\frac{s}{F^{2/3}}\right).
\label{iden1}
\end{equation}

\section{The first integral}

\label{I1}

In this Appendix we compute the first integral
\begin{equation}
I_1(T)=  \int_0^\infty  \int_0^\infty
d v\, d x\,  \left[ Z_0(x, 0 ; 0, v,T) +   Z_1(x, 0 ; 0, v,T) \right],
\label{I10}
\end{equation}
where the propagator $Z_0$ is given in Eq.~(\ref{z0}), and the Laplace transform of $Z_1$ is given in Eq.~(\ref{z1l}). We show below that $I_1(T)$ is actually a constant, independent of $T$, and is given by an amazingly simple expression
\begin{equation}
I_1(T)=C=1-\sqrt{\frac{3}{8}}.
\label{I11}
\end{equation}
Performing these integrals in closed form requires the use of a number of identities from Ref.~\cite{GR}. However, to use the available identities, we need to first express the integral in a suitable form. To summarize, the derivation is not straightforward and requires quite a few mathematical manipulations and tricks. To make it easy, we lay out the main steps used in this derivation.

\vskip 0.2cm

\noindent {\bf {Step 1:}} Upon substituting the exact $Z_0$ from Eq.~(\ref{z0}), the first integral in Eq.~(\ref{I10}) can be performed in a straightforward manner and gives
\begin{equation}
\int_0^\infty  \int_0^\infty d v\, d x\, Z_0(x, 0 ; 0, v,T)= \frac{5}{12}.
\label{I12}
\end{equation}

\vskip 0.2cm

\noindent{\bf {Step 2:}} To perform the second integral, we first take its Laplace transform with respect to $T$, substitute the result from Eq.~(\ref{z1l}), and then easily perform the integration over $x$. This yields
\begin{eqnarray}
J(s)=\int_0^{\infty}\int_0^\infty \int_0^\infty dv\, d x\, dT\, e^{-sT} Z_1(x, 0 ; 0, v,T)= -\frac{1}{2 \pi}  \int^{\infty}_{0} \int^{\infty}_{0}\int_0^{\infty} dv\,dG\, dF\nonumber \\
 \frac{(FG)^{-1/6}}{G(F+G)}\exp\left[-\frac{2}{3}\left(\frac{s^{3/2}}{F} +\frac{s^{3/2}}{G}\right)\right] \mbox{Ai}\left(\frac{s}{G^{2/3}}\right) \mbox{Ai}\left(\frac{s}{F^{2/3}}+v F^{1/3}\right).
\label{I13}
\end{eqnarray}

\vskip 0.2cm

\noindent{\bf {Step 3:}} Next, we split the term $1/\{G(F+G)\}$ in the integrand on the rhs of Eq.~(\ref{I13}) into two parts, namely,
\begin{equation}
\frac{1}{G(F+G)}= -\frac{1}{F(F+G)}+\frac{1}{FG},
\label{I14}
\end{equation}
and thus split the integral $J(s)$ into two parts, i.e., 
$J(s)=J_1(s)+J_2(s)$. This allow dealing each contribution separately, 
so that a number of identities can be used, as explained below

\vskip 0.2cm

\noindent{\bf {Step 4:}} Let us first consider the first term,
\begin{eqnarray}
J_1(s)= \frac{1}{2 \pi} \int^{\infty}_{0} \int^{\infty}_{0}\int_0^{\infty} dv\,dG\,dF\nonumber \\
 \frac{(FG)^{-1/6}}{F(F+G)}\exp\left[-\frac{2}{3}\left(\frac{s^{3/2}}{F} +\frac{s^{3/2}}{G}\right)\right] \mbox{Ai}\left(\frac{s}{G^{2/3}}\right) \mbox{Ai}\left(\frac{s}{F^{2/3}}+v F^{1/3}\right).
\label{I15}
\end{eqnarray}
Now, the integral over $G$ can be performed explicitly using the identity (\ref{iden1}). Next, for the integral over $v$, we make a change of variables: $w= F^{1/3} v + s F^{-2/3}$. Substituting back, we get
\begin{equation}
J_1(s)= \int_0^{\infty} \frac{dF}{F^{5/3}}\, \mbox{Ai}\left(\frac{s}{F^{2/3}}\right)\, \int_{sF^{-2/3}}^{\infty} \mbox{Ai}(w)\,dw.
\label{I16}
\end{equation}
Then, making a further change of variables, $sF^{-2/3}=z$, yields a simpler expression
\begin{equation}
J_1(s)= \frac{3}{2s} \int_0^{\infty} dz \mbox{Ai}(z) \int_z^{\infty} \mbox{Ai}(w) dw.
\label{I17}
\end{equation}
The rhs can be computed using integration by parts, so to give
\begin{equation}
J_1(s) = \frac{3}{4s}\left[\int_0^{\infty} \mbox{Ai}(w) dw\right]^2.
\label{I18}
\end{equation}
Using $\int_0^{\infty} \mbox{Ai}(w)dw=1/3$~\cite{AS}, finally gives the rather simple exact expression
\begin{equation}
J_1(s) = \frac{1}{12 s}.
\label{I19}
\end{equation}

\vskip 0.2cm

\noindent {\bf {Step 5:}} We now turn to the second contribution $J_2(s)$, which is given by
\begin{eqnarray}
J_2(s)= -\frac{1}{2 \pi}  \int^{\infty}_{0} \int^{\infty}_{0}\int_0^{\infty} dv\,dG\, dF\nonumber \\
 (FG)^{-7/6}\exp\left[-\frac{2}{3}\left(\frac{s^{3/2}}{F} +\frac{s^{3/2}}{G}\right)\right] \mbox{Ai}\left(\frac{s}{G^{2/3}}\right)\mbox{Ai}\left(\frac{s}{F^{2/3}}+v F^{1/3}\right).
\label{I110}
\end{eqnarray}
Now, the integral over $G$ can be separated from the integrals over $F$ and $v$, and we write
\begin{equation}
J_2(s)= -\frac{1}{2\pi}\, J_{21}(s)\, J_{22}(s),
\label{I111}
\end{equation}
where $J_{21}(s)$ involves only the integral over $G$
\begin{equation}
J_{21}(s)= \int_0^{\infty} dG\, G^{-7/6}\,\exp\left[-\frac{2}{3}\left(\frac{s^{3/2}}{G}\right)\right]\,\mbox{Ai}\left(\frac{s}{G^{2/3}}\right)
\label{I112}
\end{equation}
and $J_{22}(s)$ involves the integration over $F$ and $v$
\begin{equation}
J_{22}(s)= \int_0^{\infty} dF\, F^{-7/6}\, \exp\left[-\frac{2}{3}\left(\frac{s^{3/2}}{F}\right)\right]\,\int_0^{\infty} \mbox{Ai}(F^{1/3}v+sF^{-2/3})dv.
\label{I113}
\end{equation}

\vskip 0.2cm

\noindent {\bf {Step 6:}} We now perform the integral $J_{21}(s)$ in Eq.~(\ref{I112}). We first make a change of variables from $G$ to $y= 2s^{3/2}/{3G}$. Next, we use the identity~\cite{AS}
\begin{equation}
\mbox{Ai}\left(\left(\frac{3y}{2}\right)^{2/3}\right)= \frac{1}{\pi \sqrt{3}}\,\left(\frac{3}{2}\right)^{1/3}\, y^{1/3}\,K_{1/3}(y),
\label{iden2}
\end{equation}
where $K_{\nu}(y)$ is the modified Bessel function of index $\nu$~\cite{GR}. Substituting back in Eq.~(\ref{I112}), we get
\begin{equation}
J_{21}(s)= \frac{1}{s^{1/4}}\, \frac{1}{\pi \sqrt{2}}\,\int_0^{\infty} y^{-1/2}\,e^{-y}\,K_{1/3}(y)\, dy.
\label{I114}
\end{equation}
The integral over $y$ can be explicitly performed~\cite{GR}, and gives $\pi\sqrt{2\pi}$. Thus, we get a very simple expression
\begin{equation}
J_{21}(s)= \frac{\sqrt{\pi}}{s^{1/4}}.
\label{I115}
\end{equation}

\vskip 0.2cm

\noindent {\bf {Step 7:}} Now we turn to the computation of $J_{22}(s)$ in Eq.~(\ref{I113}). Making the change of variables from $v$ to $w=F^{1/3}v + sF^{-2/3}$, we get
\begin{equation}
J_{22}(s)= \int_0^{\infty} dF\, F^{-3/2}\, \exp\left[-\frac{2}{3}\left(\frac{s^{3/2}}{F}\right)\right]\,\int_{sF^{-2/3}}^{\infty} \mbox{Ai}(w)dw.
\label{I116}
\end{equation}
Next, we make a change of variables from $F$ to $y= 2s^{3/2}/{3F}$, so to rewrite the integral as
\begin{equation}
J_{22}(s)= \frac{1}{s^{3/4}}\, \sqrt{\frac{3}{2}}\, \int_0^{\infty} dy\, y^{-1/2}\, e^{-y}\,\int_{(3y/2)^{2/3}}^{\infty} \mbox{Ai}(w)dw .
\label{I117}
\end{equation}
Substituting the expressions for $J_{22}(s)$ and $J_{21}(s)$ in Eq.~(\ref{I111}) finally yields
\begin{equation}
J_2(s)= - \left[\sqrt{\frac{3}{8\pi}}\,A_0\right]\,\frac{1}{s}
\label{I118}
\end{equation}
where the constant $A_0$ is given by the integral
\begin{equation}
A_0= \int_0^{\infty} dy\,y^{-1/2}\, e^{-y}\,\int_{(3y/2)^{2/3}}^{\infty} \mbox{Ai}(w)dw .
\label{I119}
\end{equation}

\vskip 0.2cm

\noindent {\bf Step 8:} We have now to evaluate the constant $A_0$ in Eq.~(\ref{I119}). As a first step, let us again use the identity in Eq.~(\ref{iden2}), so to express the inner integral in Eq.~(\ref{I119}) as
\begin{equation}
\int_{(3y/2)^{2/3}}^{\infty} Ai(w) dw = \frac{1}{\pi \sqrt{3}}\int_y^{\infty} K_{1/3}(z) dz.
\label{I120}
\end{equation}
This follows by making a change of variables $w= (3z/2)^{2/3}$ on the lhs of Eq.~(\ref{I120}) and then using the identity in Eq.~(\ref{iden2}). Thus, we get
\begin{equation}
A_0= \frac{1}{\pi \sqrt{3}}\int_0^{\infty} dy\, y^{-1/2} e^{-y} \int_y^{\infty} K_{1/3}(z) dz.
\label{I121}
\end{equation}
Next, we perform integration by parts and use the incomplete Gamma function defined by $\Gamma[\alpha,y]= \int_y^{\infty} dx\, x^{\alpha-1}\, e^{-x}$. This gives
\begin{equation}
A_0= \frac{1}{\pi \sqrt{3}}\left[\sqrt{\pi}\int_0^{\infty} K_{1/3}(z)dz -\int_0^{\infty} K_{1/3}(y) \Gamma[1/2,y]\,dy\right].
\label{I122}
\end{equation}
Using the result~\cite{GR} $\int_0^{\infty} K_{1/3}(z) dz= \pi/\sqrt{3}$, we get
\begin{equation}
A_0= \frac{\sqrt{\pi}}{3}-\frac{1}{\pi \sqrt{3}}\int_0^{\infty} K_{1/3}(y) \Gamma[1/2,y]\,dy = \frac{\sqrt{\pi}}{3}-\frac{1}{\pi \sqrt{3}}\,A_1
\label{I123}
\end{equation}

\vskip 0.2cm

\noindent {\bf {Step 9:}} Concerning finally the integral $A_1$ in Eq.~(\ref{I123}), namely,
\begin{equation}
A_1= \int_0^{\infty} K_{1/3}(y) \Gamma[1/2,y]\,dy,
\label{I124}
\end{equation}
we first write
\begin{equation}
\Gamma[1/2,y]= \frac{1}{\sqrt{y}}\, e^{-y} - \frac{1}{2} \Gamma[-1/2,y],
\label{I125}
\end{equation}
which can be easily proved by integration by parts. Substituting this result in Eq.~(\ref{I124}), the first term can be easily computed, and gives$\pi\sqrt{2\pi}$. We then get
\begin{equation}
A_1= \pi\sqrt{2\pi} -\frac{1}{2} \int_0^{\infty} K_{1/3}(y)\Gamma[-1/2,y]\,dy .
\label{I126}
\end{equation}
To perform the second integral, we first use the integral representation~\cite{GR}
\begin{equation}
\Gamma[-1/2,y]= \frac{2}{\sqrt{\pi}}\,y^{-1/2}\,e^{-y}\,\int_0^{\infty} \frac{e^{-t}t^{1/2}}{y+t}\, dt.
\label{I127}
\end{equation}
Substituting this result in Eq.~(\ref{I126}) gives
\begin{equation}
A_1=  \pi\sqrt{2\pi}-\frac{1}{\sqrt{\pi}}\, \int_0^{\infty} dt\, t^{1/2}\,e^{-t}\int_0^{\infty} dy\, y^{-1/2} e^{-y} \frac{K_{1/3}(y)}{y+t}.
\label{I128}
\end{equation}
Next, we use the identity~\cite{GR}
\begin{equation}
\int_0^{\infty}  dy\, y^{-1/2}\, e^{-y}\, \frac{K_{1/3}(y)}{y+t}= 2\pi \frac{e^{t}K_{1/3}(t)}{\sqrt{t}},
\label{I129}
\end{equation}
so to perform the inner integral over $y$ in Eq.~(\ref{I128}). The integral over $t$ then becomes simple and using once again $\int_0^{\infty} K_{1/3}(t) dt= \pi/\sqrt{3}$ yields an exact expression for $A_1$, namely,
\begin{equation}
A_1= \left(\sqrt{2}-\frac{2}{\sqrt{3}}\right)\, \pi^{3/2}.
\label{I130}
\end{equation}
Substituting this result in Eq.~(\ref{I123}) gives the final expression for $A_0$
\begin{equation}
A_0= \left(1-\sqrt{\frac{2}{3}}\right)\,\sqrt{\pi}.
\label{I131}
\end{equation}

\vskip 0.2cm

\noindent {\bf {Step 10:}} Using the expression for $A_0$ in Eq.~(\ref{I118}) we compute $J_2(s)$, which, combined with the result for $J_1(s)$ in Eq.~(\ref{I19}), allows evaluating $J(s)=J_1(s)+J_2(s)$
\begin{equation}
J(s)=J_1(s)+J_2(s)=\left[\frac{7}{12}-\sqrt{\frac{3}{8}}\right]\,\frac{1}{s}.
\label{I132}
\end{equation}
This shows that the inverse Laplace transform of $J(s)$ is just the constant prefactor of $1/s$ in Eq.~(\ref{I132}). Combining this result with Eq.~(\ref{I12}) finally yields our main result
\begin{equation}
I_1(T)= C= 1-\sqrt{\frac{3}{8}}.
\label{I133}
\end{equation}

\section{The second integral}

\begin{equation}
I_2(\epsilon,t )=\int_0^\infty d x \left[Z_0(x, 0; \epsilon,0,t)+Z_1(x,0; \epsilon,0,t)\right].
\end{equation}
The first term can be easily integrated, and expanding in small $\epsilon$ yields
\begin{equation}
\int_0^\infty d x Z_0(x,0;\epsilon,0,t)=\frac{1}{4\sqrt{\pi t}}-\frac{\sqrt{3}}{2 \pi t^2}\epsilon + \ldots
\label{contribution0}
\end{equation}
We next take the Laplace transform of the second term with respect to $t$: 
$\tilde{Z}_1(x,0;\epsilon,0,s)=\int_0^{\infty} 
Z_1(x,0;\epsilon,0,t)\,e^{-st}\, dt$
and then integrate over $x$ using the expression in Eq. (\ref{z1l}).
To extract the leading $\epsilon$ dependence, it turns out to be
useful to split $e^{-F\epsilon}=e^{-F\epsilon}-1+1$. This gives
\begin{equation}
\int_0^\infty d x \tilde{Z}_1(x,0;\epsilon,0,s)=\int_{0}^{\infty}\Phi(F,s)dF + \int_{0}^{\infty}dF\left(e^{-F\epsilon}-1 \right) \Phi(F,s),
\label{contribution1}
\end{equation}
where
\begin{eqnarray}
\Phi(F,s)=-\frac{1}{2 \pi F^{1/6}}\int_{0}^{\infty}\frac{dG}{G^{7/6}(F+G)}\times \nonumber \\
\exp\left[-\frac{2}{3}\left(\frac{s^{3/2}}{F} +\frac{s^{3/2}}{G}\right)\right] \mbox{Ai}\left(\frac{s}{G^{2/3}}\right)\mbox{Ai}\left(\frac{s}{F^{2/3}}\right).
\label{Phi}
\end{eqnarray}
The integration over $F$ in the first term of Eq. (\ref{contribution1}) can be carried out by resorting to the identity (\ref{iden1}), so to get
\begin{equation}
\int_{0}^{\infty}\Phi(F,s)dF = - \int_{0}^{\infty} \frac{dG}{G^{4/3}} \mbox{Ai}^2\left(\frac{s}{G^{2/3}}\right)=-\frac{1}{4\sqrt{s}}.
\end{equation}
This contribution, after performing the inverse Laplace transform, cancels the one coming from Eq. (\ref{contribution0}) (at the leading order in the small $\epsilon$ expansion).

We address now the second term in Eq. (\ref{contribution1}) in the small $\epsilon$ expansion. By means of the change of variables $\epsilon F = z$, the leading order of the second term reads
\begin{equation}
\int_{0}^{\infty}dF\left(e^{-F\epsilon}-1 \right) \Phi(F,s)=
\frac{\epsilon^{1/6}\mbox{Ai}\left(0\right)}{2\pi} B_1 B_2(s),
\end{equation}
where
\begin{equation}
B_1=\int_{0}^{\infty} dz \frac{\left( 1-e^{-z}\right) }{z^{7/6}}=-\Gamma(-1/6)
\end{equation}
and
\begin{equation}
B_2(s)=\int_{0}^{\infty} \frac{dG}{G^{7/6}}\, 
\exp\left(-\frac{2}{3}\frac{s^{3/2}}{G} \right)\,
\mbox{Ai}\left(\frac{s}{G^{2/3}}\right).
\end{equation}
Setting $y=s^{3/2}/G$, we get
\begin{equation}
B_2(s)=\frac{1}{s^{1/4}}\int_{0}^{\infty} 
\frac{dy}{y^{5/6}}\,\exp\left(-\frac{2}{3}y\right)\, \mbox{Ai}\left( 
y^{2/3}\right) = \frac{\sqrt{\pi}}{s^{1/4}}.
\end{equation}
Using $\mbox{Ai}\left( 0\right)=1/(3^{2/3} \Gamma(2/3))$, we finally get
\begin{equation}
\int_{0}^{\infty}dF\left(e^{-F\epsilon}-1 \right) \Phi(F,s)=\frac{3^{5/6}\Gamma(2/3)}{2^{2/3}\pi}\frac{\epsilon^{1/6}}{s^{1/4}}.
\end{equation}
Then, by inverting the Laplace transform, we obtain the leading 
order contribution
\begin{equation}
I_2(\epsilon,t )= 
\frac{3^{5/6}\Gamma(2/3)}{2^{2/3}\pi \Gamma(1/4)} 
\frac{\epsilon^{1/6}}{t^{3/4}}+{\cal O}(\epsilon^{1/3}).
\label{I2final}
\end{equation}

\section{The third integral}

Remark that the integral
\begin{equation}
I_3(\epsilon,t )=\int_{-\infty}^{+\infty} d v \int_0^\infty  d x \left[Z_0(x, v; \epsilon,0,t)+Z_1(x,v; \epsilon,0,t)\right]
\end{equation}
represents the survival probability at time $t$ of the process 
started at the origin with vanishing velocity. The Laplace transform of this 
quantity, with respect to $t$, has been computed in \cite{Burkhardt}, and 
reads
\begin{equation}
\tilde{I}_3(\epsilon,s )=\frac{1}{s} -\int_{0}^{\infty}dF \Psi(F,s) -\int_{0}^{\infty}dF (e^{-\epsilon F}-1) \Psi(F,s),
\end{equation}
where
\begin{equation}
\Psi(F,s)=\frac{\mbox{Ai}\left( s/F^{2/3}\right)}{F^{5/3}}\left[ 1 + \frac{\Gamma(-\frac{1}{2},\frac{2}{3}\frac{s^{3/2}}{F})}{4\sqrt{\pi}}\right].
\end{equation}
For the second term, we can set $s/F^{2/3}=y$, so to make the dependence on $s$ explicit, and get after some algebra
\begin{equation}
\int_{0}^{\infty}dF \Psi(F,s)= \frac{3}{2s} \int_{0}^{\infty}dy \mbox{Ai}\left(y \right)\left[ 1 + \frac{\Gamma(-\frac{1}{2},\frac{2}{3}y^{3/2})}{4\sqrt{\pi}}\right]=\frac{1}{s}.
\end{equation}
This contribution cancels the $\frac{1}{s}$ of the first term. The third term can be computed by resorting to the substitution $z=\epsilon F$, namely
\begin{equation}
\int_{0}^{\infty}dF (e^{-\epsilon F}-1) \Psi(F,s)=\epsilon^{2/3} \mbox{Ai}(0) \int_{0}^{\infty} \frac{dz}{z^{5/3}}\left(e^{-z}-1 \right) \left[ 1+ \frac{\Gamma(-\frac{1}{2},\frac{2}{3}\frac{\epsilon s^{3/2}}{z})}{4\sqrt{\pi}}\right]. 
\end{equation}
Then, by noting that, for small $x$, $\Gamma(-1/2,x) \to 2/\sqrt{x}$, we get
\begin{equation}
\int_{0}^{\infty}dF (e^{-\epsilon F}-1) \Psi(F,s)=\sqrt{\frac{3}{8\pi}}\mbox{Ai}(0) \frac{\epsilon^{1/6}}{s^{3/4}}\int_{0}^{\infty}\frac{dz}{z^{7/6}}\left(1-e^{-z} \right)=\frac{2^{5/6}\Gamma(-4/3)}{3^{2/3}\pi}\frac{\epsilon^{1/6}}{s^{3/4}}. 
\end{equation}
By finally performing the inverse Laplace transform, we obtain
\begin{equation}
I_3(\epsilon,t)=  \frac{2^{5/6}\Gamma(-4/3)}{3^{2/3}\pi\Gamma(3/4)} \frac{\epsilon^{1/6}}{t^{1/4}}+O(\epsilon^{1/3}).
\label{I3}
\end{equation}


\section*{References}
\begin{thebibliography}{10}

\bibitem{Levy} Levy, P.: Sur Certains Processus Stochastiques Homog\'enes. Comp. Math. {\bf 7}, 283 (1939)

\bibitem{Feller} Feller, W.: An Introduction to Probability Theory and its Applications. New York, Wiley (1968)

\bibitem{Yor} Majumdar, S.~N., Randon-Furling, J., Kearney, M.~J., Yor, M.: On The Time To Reach Maximum For A Variety Of Constrained Brownian Motions. J. Phys. A: Math. Theor. {\bf 41}, 365005 (2008)

\bibitem{RM} Randon-Furling, J., Majumdar. S.~N.: Distribution of the time at which the deviation of a Brownian motion is maximum before its first-passage time. J. Stat. Mech. P10008 (2007)

\bibitem{MB} Majumdar, S.~N., Bouchaud, J.~-P.: Optimal Time to Sell a Stock in Black-Scholes Model. Quantitative Finance {\bf 8}, 753 (2008)

\bibitem{SL} Schehr, G., Le Doussal, P.: Extreme value statistics from the Real Space Renormalization Group: Brownian Motion, Bessel Processes and Continuous Time Random Walks. arXiv:0910:4913

\bibitem{hull1} Randon-Furling, J., Majumdar, S.~N., Comtet, A.: Convex Hull of N planar Brownian Motions: Application to Ecology.  Phys. Rev. Lett. {\bf 103}, 140602 (2009)

\bibitem{hull2} Majumdar, S.~N., Comtet, A., Randon-Furling, J.: Random Convex Hulls and Extreme Value Statistics. J. Stat. Phys., in press, arXiv: 0912:0631

\bibitem{MRZ} Majumdar, S.~N., Rosso, A., Zoia, A.: Hitting Probability for Anomalous Diffusion Processes. Phys. Rev. Lett., in press, arXiv:0911:3815

\bibitem{Burkhardt} Burkhardt, T.~W.: Semiflexible Polymer in the Half Plane and Statistics of the Integral of a Brownian Curve. J. Phys. A: Math. Gen. {\bf 26}, L1157 (1993)

\bibitem{MB1} Majumdar, S.~N., Bray, A.~J.: Spatial Persistence of Fluctuating Interfaces. Phys. Rev. Lett. {\bf 86}, 3700 (2001)

\bibitem{GBDT} Golubovic, L., Bruinsma, R.: Surface diffusion and fluctuations of growing interfaces. Phys. Rev. Lett. {\bf 66}, 321 (1991); Das Sarma, S., Tamborenea, P.: A new universality class for kinetic growth: One-dimensional molecular-beam epitaxy. Phys. Rev. Lett. {\bf 66} 325 (1991)

\bibitem{PV} Valageas, P.: Statistical properties of the Burgers equation with Brownian initial velocity. J. Stat. Phys. {\bf 134}, 589 (2009)

\bibitem{McKean} McKean, H.~P.: A winding problem for a resonator driven by a white noise. J. Math. Kyoto Univ. {\bf 2} 227 (1963)

\bibitem{Goldman} Goldman, M.: On the first passage of the integrated Wiener process. Ann. Math. Stat. {\bf 42} 2150 (1971)

\bibitem{MW} Marshall, T.~W., Watson, E.~J.: A drop of ink falls from my pen... it comes to earth, I know not when. J. Phys. A: Math. Gen. {\bf 18}, 3531 (1985)

\bibitem{Sinai} Sinai, Ya.~G.: Distribution of some functionals of the integral of a random walk. Theor. Math. Phys. {\bf 90}, 219 (1992)

\bibitem{Lachal} Lachal, A.: Last passage time for integrated Brownian motion. Stoch. Proc. Appl. {\bf 49}, 57 (1994)

\bibitem{MB2} Majumdar, S.~N., Bray, A.~J.: Persistence With Partial Survival. Phys. Rev. Lett. {\bf 81}, 1142 (1998)

\bibitem{GL} De Smedt, G., Godreche, G., Luck, J.~M.: Partial survival and inelastic collapse for a randomly accelerated particle. Europhys. Lett. {\bf 53}, 438 (2001)

\bibitem{Burkhardt2} Burkhardt, T.~W.: Dynamics of Absorption of a Randomly Accelerated Particle. J. Phys. A: Math. Gen. {\bf 33}, L429 (2000)

\bibitem{Gyorgy} Gyoergyi. G., Moloney, N.~R., Ozogany, K., Racz, Z.: Maximal height statistics for 1/f$^{\alpha}$ signals. Phys. Rev. E {\bf 75}, 021123 (2007)

\bibitem{Burkhardt3} Burkhardt, T.~W.: First-Passage and Extreme-Value Statistics of a Particle Subject to a Constant Force Plus a Random Force. J. Stat. Phys. {\bf 133}, 217 (2008)

\bibitem{GR} Gradshteyn, I.~S., Ryzhik, I.~M.: Tables of Integrals, Series, and Products. Academic, New York (1980)


\bibitem{Airy} Majumdar, S.~N., Comtet, A.: Airy Distribution Function: From the Area under a Brownian Excursion to the Maximal Height of Fluctuating Interfaces. J. Stat. Phys. {\bf 119}, 777 (2005)

\bibitem{foot} Note that for a symmetric process, the distribution of
the location of its minimum coincides with that of its maximum.

\bibitem{AS} Abramowitz, A., Stegun, I.~A.: Handbook of Mathematical Functions. Dover, New York (1965)



\end{thebibliography}
\end{document}